# MatterGPT: A Generative Transformer for Multi-Property Inverse Design of Solid-State Materials

Yan Chen[1*], Xueru Wang[1], Xiaobin Deng[2], Yilun Liu[1], Xi Chen[2], Yunwei Zhang[3], Lei Wang[4,5*], Hang Xiao[2*]

[1] *Laboratory for multiscale mechanics and medical science, SV LAB, School of Aerospace, Xi'an Jiaotong University, Xi'an 710049, China*

[2] *School of Interdisciplinary Studies, Lingnan University, Tuen Mun, Hong Kong SAR, China*

[3] *State Key Laboratory of Optoelectronic Materials and Technologies, School of Physics, Sun Yat-Sen University, Guangzhou 510275, China*

[4] *National Laboratory of Solid-State Microstructures, School of Physics, Nanjing University, Nanjing 210093, China*

[5] *Collaborative Innovation Center of Advanced Microstructures, Nanjing University, Nanjing 210093, China*

\* Corresponding authors.

E-mail: yanchen@xjtu.edu.cn; leiwang@nju.edu.cn; hangxiao@ln.edu.hk.

**ABSTRACT**

Inverse design of solid-state materials with desired properties represents a formidable challenge in materials science. Although recent generative models have demonstrated potential, their adoption has been hindered by limitations such as inefficiency, architectural constraints and restricted open-source availability. The representation of crystal structures using the SLICES (Simplified Line-Input Crystal-Encoding System) notation as a string of characters enables the use of state-of-the-art natural language processing models, such as Transformers, for crystal design. Drawing inspiration from the success of GPT models in generating coherent text, we trained a generative Transformer on the next-token prediction task to generate solid-state materials with targeted properties. We demonstrate MatterGPT's capability to generate de novo crystal structures with targeted single properties, including both lattice-insensitive (formation energy) and lattice-sensitive (band gap) properties. Furthermore, we extend MatterGPT to simultaneously target multiple properties, addressing the complex challenge of multi-objective inverse design of crystals. Our approach showcases high validity, uniqueness, and novelty in generated structures, as well as the ability to generate materials with properties beyond the training data distribution. This work represents a significant step forward in computational materials discovery, offering a powerful and open tool for designing materials with tailored properties for various applications in energy, electronics, and beyond.

# Introduction

The holy grail of materials science is to engineer materials with desired properties – a concept known as inverse design [1]. This challenge is particularly acute for crystalline materials, where the chemical space of possible structures has been estimated at $10^{100}$ [2], many orders of magnitude larger than that of drug-like molecules [3]. Traditional high-throughput screening methods, constrained by existing databases, struggle to efficiently explore the vast chemical space of potential crystal structures [4]. Crystal structure prediction (CSP) has emerged as a promising approach for exploring beyond the chemical space of known materials. However, CSP is hindered by high computational costs and the stochastic nature of many global optimization algorithms used in CSP can lead to incomplete or biased exploration of the configurational space [5].

In recent years, deep generative models have emerged as powerful tools for inorganic material inverse design, with various architectures like Variational Autoencoders (VAEs) [6], Generative Adversarial Networks (GANs) [7], and diffusion models[8] leading the way. Early approaches utilizing 3D image-based or voxel-based crystal representations faced challenges in scalability and accuracy. For instance, Noh et al. proposed iMatGen, a VAE-based model for inverse generation of V$x$O$y$ crystals [9]. Long et al. developed CCDCGAN, a GAN-based model for generating novel binary Bi-Se crystal structures [10]. Court et al. introduced Cond-DFC-VAE, capable of generating novel crystal structures by sampling from the VAE latent space [11]. However, these methods were largely limited to specific structural systems, often hampered by limitations such as high computational costs, memory intensity, and were prone to generating invalid structures.

To address these limitations, researchers have explored alternative representations. Using crystal graph representation, Xie et al. proposed CDVAE [12], a crystal diffusion VAE framework that directly reconstructs stable crystals by refining atom types, coordinates and the periodic lattice in a diffusion process. Building on this, Jiao et al. introduced DiffCSP [13], which conducts joint diffusion on lattices and fractional coordinates. More recently, Microsoft Research AI4Science presented MatterGen [14], an advancement over CDVAE, capable of generating stable and novel crystals with on-demand properties. However, diffusion-based models must specially handle discrete atomic identities and often suffer from low training and sampling speeds [15]. Moreover, MatterGen is not open source, restricting its availability to the broader materials science community. Alternative approaches include the invertible Fourier-transformed crystal properties (FTCP) representation by Ren et al. [16] and the WyCryst model by Zhu et al. [15], which incorporates Wyckoff-based feature representation. While these models can inversely design unique crystals with user-defined properties by sampling a property-structured latent space from a VAE, they are limited in the number of crystals they can generate and have relatively low success rates. Despite these advancements, there remains a lack of practical frameworks for crystal inverse design that can overcome the limitations of computational cost, generalization across different crystal systems, and high success rates in generating valid structures with desired properties.

Transformer-based models, initially breakthrough in natural language processing [17], have rapidly evolved to reshape various domains, including bioinformatics, chemistry and materials science, etc [18-22]. In molecular design, these models, leveraging representations like SMILES [23], have revolutionized structure generation and optimization [24]. Transformer-based approaches have established a new paradigm for material design, evidenced by models such as MolGPT [25], DNAGPT [26], ProteinGPT [27], among others. However, applying Transformers to crystal structure generation remains challenging, primarily due to the limitations of existing crystal representations. Many methods rely on conventional formats like Crystallographic Information Files (CIF), which suffer from representational redundancy and lack of invariance. For instance, CrystaLLM generates CIF syntax for inorganic compounds [28], while Gruver et al. fine-tuned the LLaMA-2 70B model for materials generation by incorporating string representations of crystal structure information with task-specific text prompts [29]. The fundamental issue with these approaches is their sensitivity to Euclidean transformations: the same crystal structure can have countless CIF representations due to translation and rotation operations. This redundancy significantly hampers the efficiency of Transformer-based models, which must learn multiple representations of identical structures. In addition, large language models (LLMs) models have also been used to generate chemically valid material compositions [30, 31], but they often struggle to capture the complex spatial information crucial for material properties. Therefore, there is a clear need for a more sophisticated crystal representation that is both invertible and invariant to Euclidean transformations.

To address these limitations, in our previous work, we developed the Simplified Line-Input Crystal-Encoding System (SLICES) [32], a novel approach to crystal representation. SLICES offers an invertible and invariant "crystal language," analogous to SMILES for molecules, enabling efficient encoding of crystal structures into simple strings. Building upon this foundation, we now introduce MatterGPT—a pre-trained generative Transformer-decoder model designed for on-demand, de novo generation of solid-state materials. By integrating LLMs with the SLICES representation, MatterGPT seeks to overcome existing limitations in crystal structure generation. MatterGPT employs a ChatGPT-like autoregressive decoder-only Transformer architecture to learn the grammar of SLICES and the intrinsic relationships between crystal structures and their properties. We demonstrate MatterGPT's capability to generate crystal structures with desired single properties, such as formation energy and band gap, and its capability to target multiple properties simultaneously.

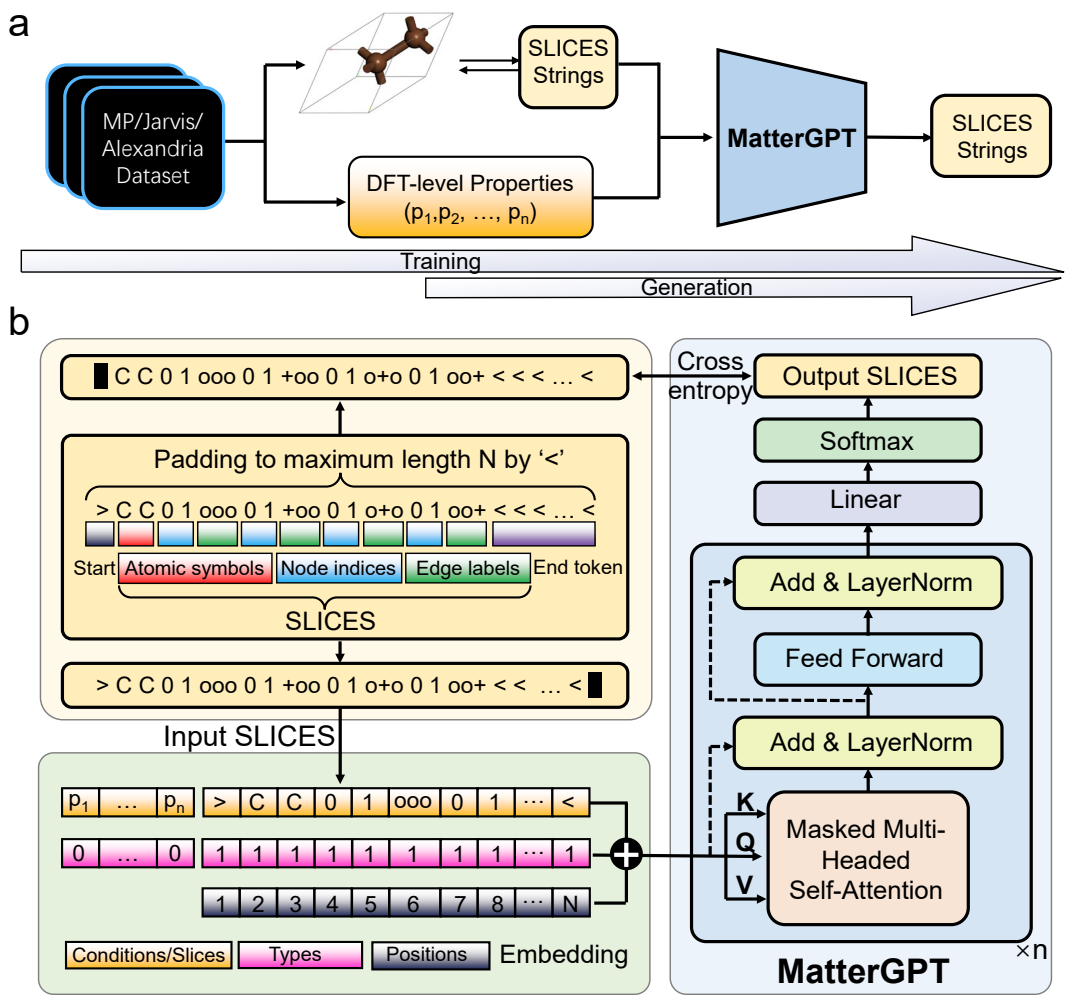

**Figure 1. The architecture of MatterGPT**. **a,** pipeline for training and sampling using the conditional MatterGPT model. **b,** the preprocessing of SLICES strings before inputting them into the autoregressive decoder-only Transformer architecture with masked multi-head self-attention module, which involves adding a start token '>' at the beginning of each SLICES string and padding the end with the token '<' to reach a maximum length of $N$. The SLICES string and targeted properties are first embedded and then concatenated, with type embeddings employed to distinguish between them. Trainable position embeddings are used to capture the order of tokens.

## RESULTS AND DISCUSSIONS

### Conditional Generation Based on Single Properties

The Alex-20 dataset, derived from the Alexandria database [33], includes materials with fewer than 20 atoms per unit cell. After filtering out crystals with atomic number exceeding 86, low-dimensional structures (0D, 1D, 2D), and metallic structures to avoid biasing, the final dataset comprises 280,033 unique crystal structures. (see **MATERIALS AND METHODS** for further details).

Before sampling novel SLICES, we first train MatterGPT model to learn the grammar of SLICES and the intrinsic relationships between them and their corresponding properties, as illustrated in **Figure 1**(a). The crystal language SLICES consist of three components: atomic symbols, node indices and edge labels[32]. In this work, we consider 83 atomic symbols, 20 node indices (integers from 0 to 19) and 27 edge labels composed of 3 characters (i.e., '+', '-' and 'o'). In the original SLICES paper, three separate tokens were used for individual edge labels [32]. We improved this method by consolidating these tokens into a single one, significantly reducing sequence length and enhancing the efficiency of the learning process. In MatterGPT model, we employ a vocabulary of 132 tokens in total, including 130 tokens for SLICES, plus 2 special tokens: a start token '>' and padding token '<'. A start token '>' is added at the beginning of each SLICES string, and all SLICES strings are padded with the padding token '<' to maximum length $N$.

Here, MatterGPT adopts a ChatGPT-like autoregressive decoder-only Transformer architecture with masked multi-head self-attention module [17]. The model was trained with a causal language modeling (CLM) objective, i.e., predicting the next token in a sequence to sample intact and syntactically coherent SLICES string meeting the desired properties. The detailed architecture of MatterGPT is illustrated in **Figure 1b**, and all training hyperparameters are provided in **Table S1**. During generation, the start token '>' is first fed into MatterGPT model and outputs the next token sequence-by-sequence. This process is repeated iteratively during sampling until the full SLICES strings are produced up to maximum length $N$, as shown in **Figure 1a**.

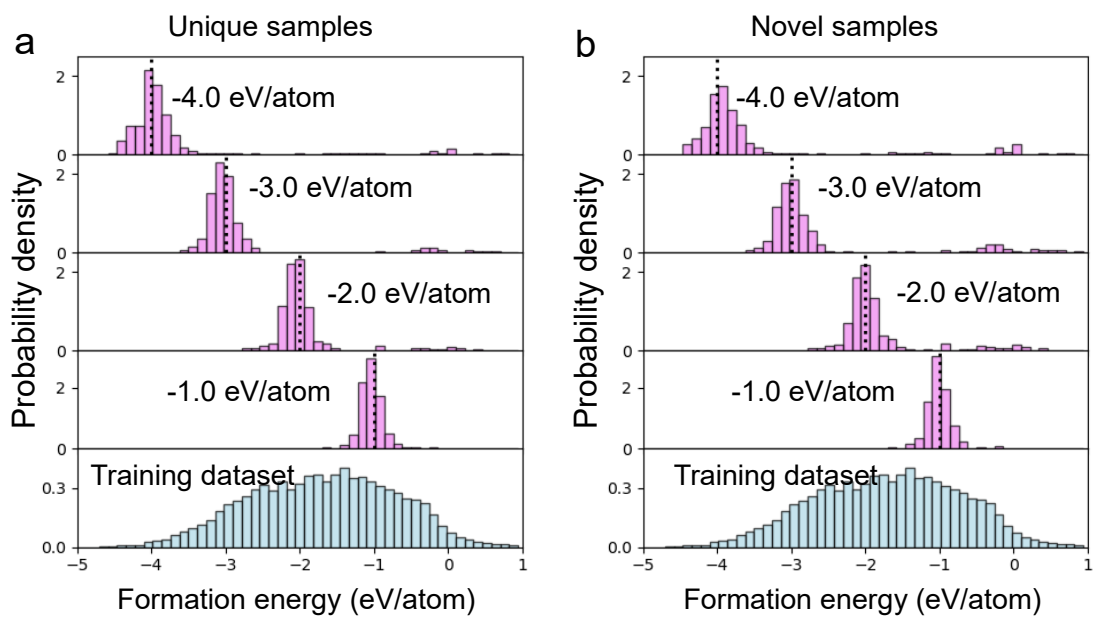

**Figure 2. De novo crystal generation by MatterGPT targeting specific formation energies (-1.0, -2.0, -3.0, -4.0 eV/atom). a,** unique samples and **b,** novel samples. The histogram at the bottom shows the distributions of formation energies per atom in the Alex-20

dataset.

For crystalline materials, their properties are entirely encoded within their periodic unit cells[34]. However, generative models often generate crystal structures with certain lattice errors, leading to inaccuracies in the learned structure-property relationships[35]. Not all properties of crystalline structures are strictly dependent on lattice constants; some are relatively insensitive to these parameters. Thus, these properties can be divided into two categories. The first includes those that are highly sensitive to precise lattice constants and structural parameters, such as electronic band structures or phonon frequencies, which can be significantly affected by minor deviations in lattice dimensions. The second category comprises properties that are stable across a range of lattice parameters and are governed more by overall symmetry and composition rather than precise lattice dimensions. Examples include formation energy, certain mechanical properties like bulk modulus, and some optical properties.

Table 1 Generative Performance of MatterGPT for Targeted Properties.

| Condition | Target | Validity (%) | Uniqueness (%) | Novelty (%) | MAPE |
|---|---|---|---|---|---|
| Formation energy (eV/atom) | -1.0 | 92.6 | 94.2 | 58.3 | 12.6 |
| | -2.0 | 93.6 | 95.5 | 58.8 | 11.9 |
| | -3.0 | 92.0 | 96.3 | 56.2 | 12.2 |
| | -4.0 | 95.0 | 89.9 | 57.4 | 10.9 |
| Energy band gap (eV) | 1.0 | 94.4 | 97.2 | 41.0 | 51.1 |
| | 2.0 | 96.8 | 97.7 | 39.7 | 33.2 |
| | 3.0 | 95.0 | 97.3 | 40.7 | 31.4 |
| | 4.0 | 93.2 | 96.8 | 40.6 | 32.5 |

First, we consider the lattice-insensitive property, formation energy ($E_f$), as a starting point, which serves as one of the primary indicators of the stability of a crystal structure[10, 11, 36]. During training, the embeddings of targeted properties are concatenated with the SLICES embedding, with type embeddings used to distinguish between properties and SLICES. The position embeddings are trainable and capture the order of tokens, similar to the approach used in the original Transformer model[17]. (See **MATERIALS AND METHODS** for train details and **Table S1** for hyperparameters). During sampling, we employed MatterGPT to generate 500 crystals targeting specific $E_f$, i.e., -1.0, -2.0, -3.0 and -4.0 eV/atom, respectively. The generated SLICES are then reconstructed into crystal structures and their validity, uniqueness and novelty are calculated to assess the generative performance of MatterGPT for targeted properties, as shown in **Table 1** (see **MATERIALS AND METHODS** for detailed evaluation criteria). The validity of the generated SLICES all exceeds 90%, indicating that MatterGPT model effectively captures the grammatical structure of SLICES. Additionally, the uniqueness of generated SLICES is found to be near or above 90%, indicating that MatterGPT model tends to generate unique SLICES. Furthermore, even after excluding SLICES present in training dataset, more than half of the unique SLICES remain. Note that, in all the above evaluation processes, around 4% of SLICES that could not be converted back to crystal structures were removed.

Next, we follow the workflow of Materials Project (MP) database to calculate formation energy $E_f$ at the Perdew-Burke-Ernzerhof (PBE) level of crystals generated by MatterGPT. As shown in **Figure 2a**, MatterGPT successfully samples crystals with targeted $E_f$ for unique SLICES. Notably, when targeting an $E_f$ of -4.0 eV/atom, MatterGPT generates crystals predominantly clustered around the target value, despite the training dataset containing limited data near -4 eV/atom. The mean absolute percentage error (MAPE) of the generated crystals for various targets is slightly above 10%, indicating the model's accuracy in generating outputs aligned with target properties. In addition, for novel SLICES, the distribution of $E_f$ also centers around the target values (**Figure 2b**). The similarity between these two distributions suggests that MatterGPT has effectively learned the intrinsic relationship between crystal structures and formation energies, enabling it to generate novel crystal structures that adhere to the same structure-formation energy relationship rules in the training dataset.

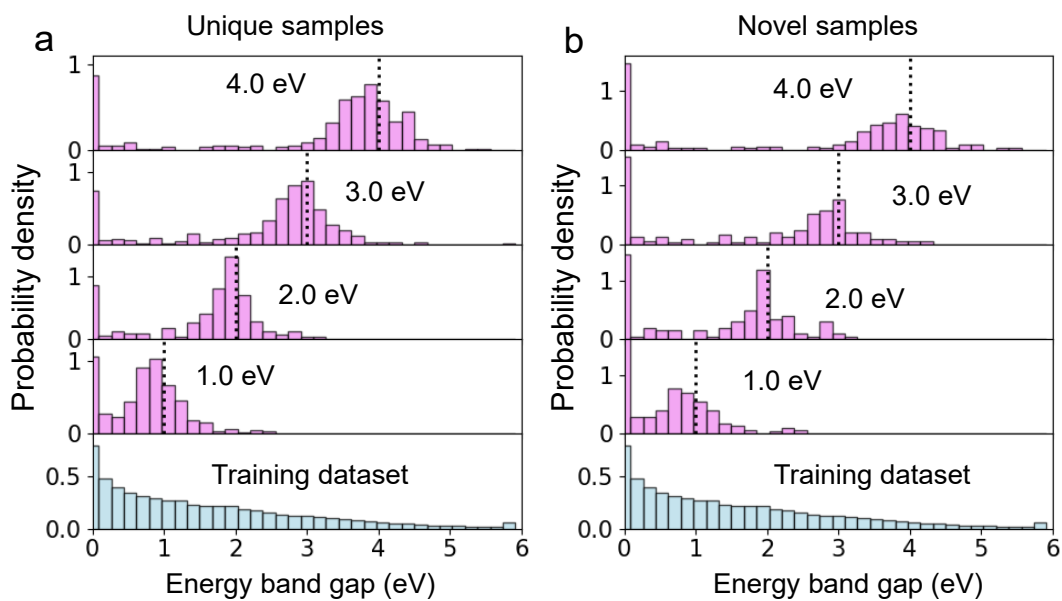

**Figure 3. De novo crystal generation by MatterGPT targeting specific band gap $E_g$=-1.0, -2.0, -3.0, -4.0 eV/atom. a,** unique samples and **b,** novel samples. The bottom histogram shows the distributions of energy band gap in the Alex-20 dataset.

To further demonstrate the capabilities of MatterGPT for on-demand crystal generation,

we train another MatterGPT model targeting a lattice-sensitive property, energy band gap $E_g$, which is crucial for understanding the electronic properties of materials and their potential applications in semiconductors, photovoltaics, hydrogen production, storage materials, etc[37, 38]. We leverage the trained MatterGPT model for $E_g$ to generate 500 crystals with a desired energy band gap of 1.0, 2.0, 3.0 and 4.0 eV, respectively. As shown in **Table 1**, the validity and uniqueness of the generated SLICES both exceed 90%, while the novelty ratio is approximately 40%, significantly lower than that observed for formation energy. To obtain $E_g$ of the generated crystals, we follow the workflow of the MP database to calculate PBE-level $E_g$ (see **MATERIALS AND METHODS**). As shown in **Figure 3**, the generated crystals are well distributed around the targeted values. In addition, the MAPE for the generated crystals is slightly above 30% for target $E_g$ of 2.0, 3.0 and 4.0 eV, but around 50% for $E_g$ of 1.0 eV, indicating that the generative model's directional accuracy is lower for target near the metal-insulator transition. The results also suggest that on-demand design of crystals targeting lattice-sensitive properties is considerably more challenging than for lattice-insensitive properties. However, this outcome highlights the effectiveness of our inverse design approach, especially given the sensitivity of the band gap ($E_g$) of bulk crystals to lattice perturbations, which typically makes capturing the intrinsic relationship between band gaps and crystal structures a challenging task [39, 40]. Moreover, considering the distribution bias of the training datasets, where the band gap values become more concentrated as they approach zero, [41] the proposed MatterGPT model effectively captures the intrinsic lattice information and learns the relationship between crystal structure and band gap.

In **Figure 4**, we present the top 3 novel de novo generated crystals with band gaps closest to the target values of $E_g$=1.0, 2.0, 3.0, and 4.0 eV, respectively. The band structures of these materials are depicted in **Figure 5**. The presented 12 crystals represent a variety of crystal systems, including hexagonal, orthorhombic, monoclinic and triclinic, and encompass a total of 31 unique elements. Notably, this approach can also be extended to generate novel crystals with specific crystal systems or space groups for solid-state materials [15], i.e., incorporating crystal symmetry information into MatterGPT model as an additional condition [29]. The results further demonstrate that MatterGPT model effectively learns the comprehensive information provided by training dataset, thereby generating novel crystal structures without special constraints on space group, element type, or other factors.

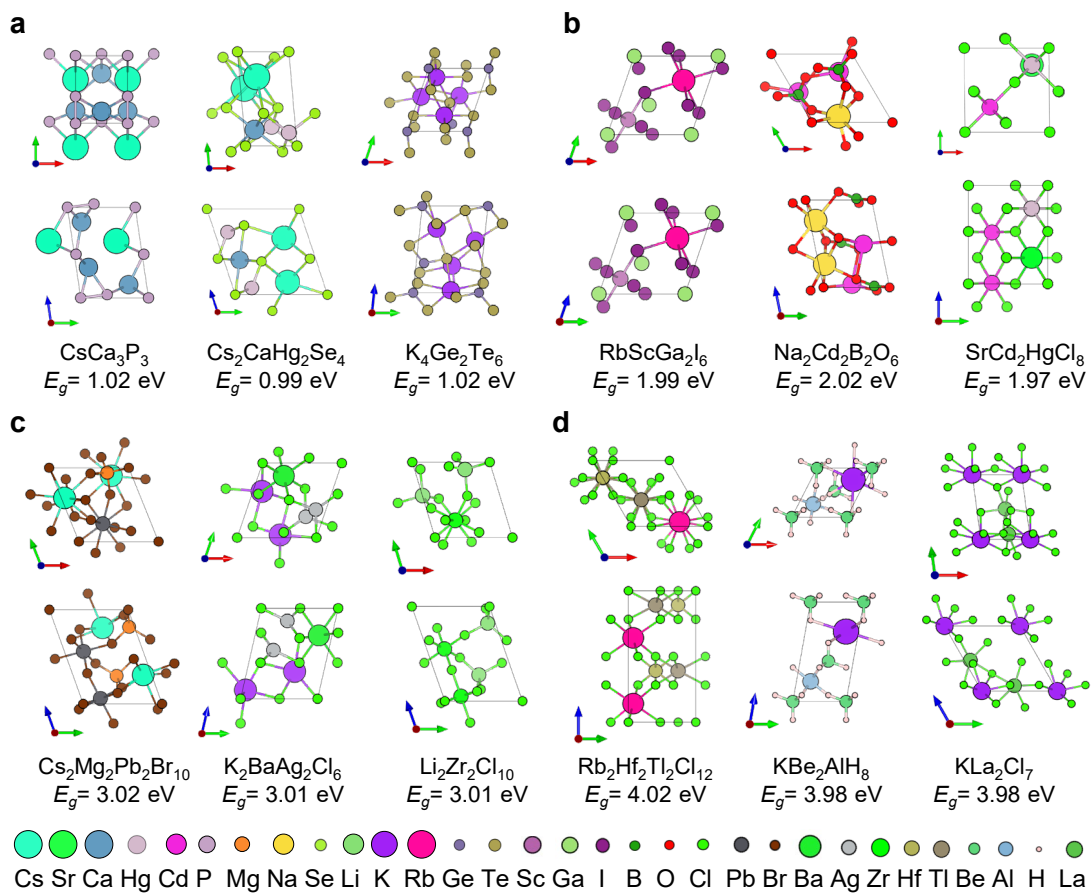

**Figure 4. Top and side views of 3 novel de novo generated crystals with targeted band gaps. a-d, $E_g$=1.0, 2.0, 3.0, 4.0 eV, respectively.** For each generated crystal, the value of energy band gap at the PBE level is provided below.

The success of MatterGPT stems from the effectiveness of the SLICES representation, which accurately encodes critical structural information with a 94.95% invertible rate. This high fidelity in representing crystal structures enables MatterGPT to effectively learn and navigate the complex relationship between crystal structures and their properties. Building on this capability, the following section will demonstrate MatterGPT's ability to perform inverse design targeting multiple properties simultaneously.

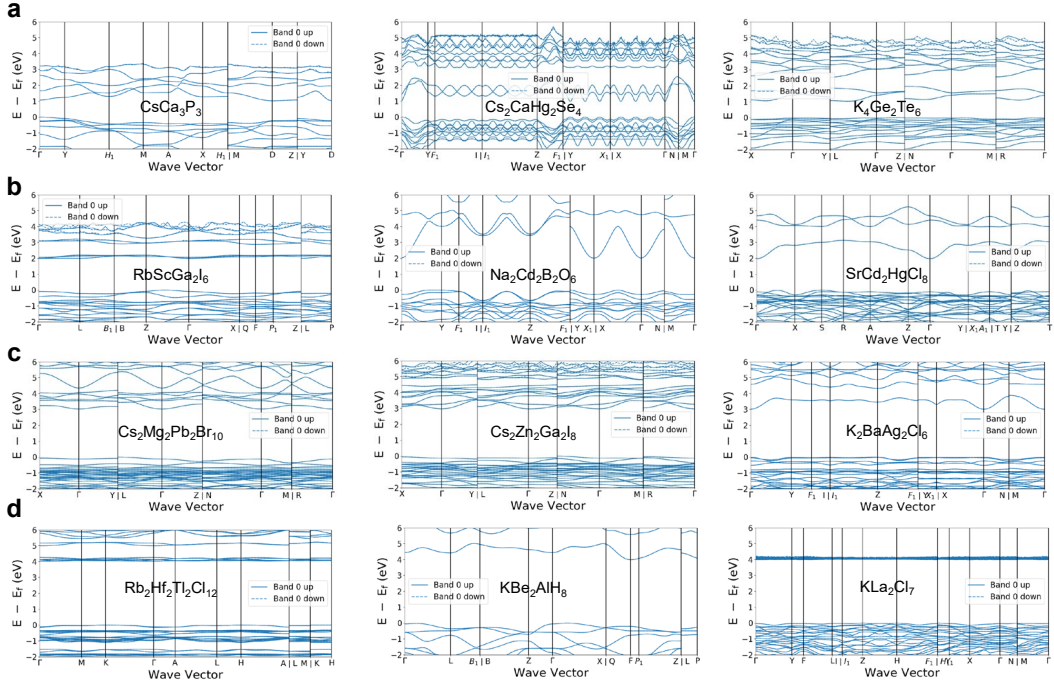

**Figure 5.** Energy band structures at the PBE level of de novo generated crystals in **Figure 4**. a-d, $E_g$=1.0, 2.0, 3.0, 4.0 eV, respectively.

## Conditional Generation Based on Multiple Properties

In practical material design, meeting multiple property requirements simultaneously is crucial.[38] For example, thermoelectric materials must balance high thermoelectric figure of merit with good electrical conductivity and low thermal conductivity [42]. Magnetic materials require high magnetic density and a low Herfindahl–Hirschman Index (HHI) [14], while ultra-incompressible crystals must maintain stability [43]. Achieving these multi-property objectives is particularly challenging for crystals due to their vast design space, numerous elemental combinations, and complex structure-property relationships.

To address this challenge, we demonstrate MatterGPT's capability to generate de novo crystals tailored to multiple properties, using band gap ($E_g$) and formation energy ($E_f$) as representative examples. We trained MatterGPT on the Alex-20 dataset, concatenating the embeddings of $E_g$ and $E_f$ with the SLICES embedding (**Figure 1b**). The model architecture remains consistent with the single-property version.

To showcase MatterGPT's performance, we targeted two property pairs: (-1.0 eV/atom, 2.0 eV) and (-4.0 eV/atom, 4.0 eV) for ($E_g$, $E_f$). The latter pair, being at the edge of the training dataset (**Figure 6a**), demonstrates the model's out-of-distribution generative capability. As shown in **Figures 6b-c**, MatterGPT effectively samples crystals centered around these desired property pairs. Note that, we only present novel crystals, as the distribution difference between unique and novel samples is negligible. **Figures 6d-e** present the top 3 novel de novo generated crystals with properties closest to the target values, which also encompasses a wide variety of element types and space group types. We also developed a webapp for inference using trained MatterGPT models targeting desired ($E_g$, $E_f$), allowing users to directly generate SLICES based on their specifications and convert them into crystal files (see 'GUI Demo' in **MATERIALS AND METHODS**).

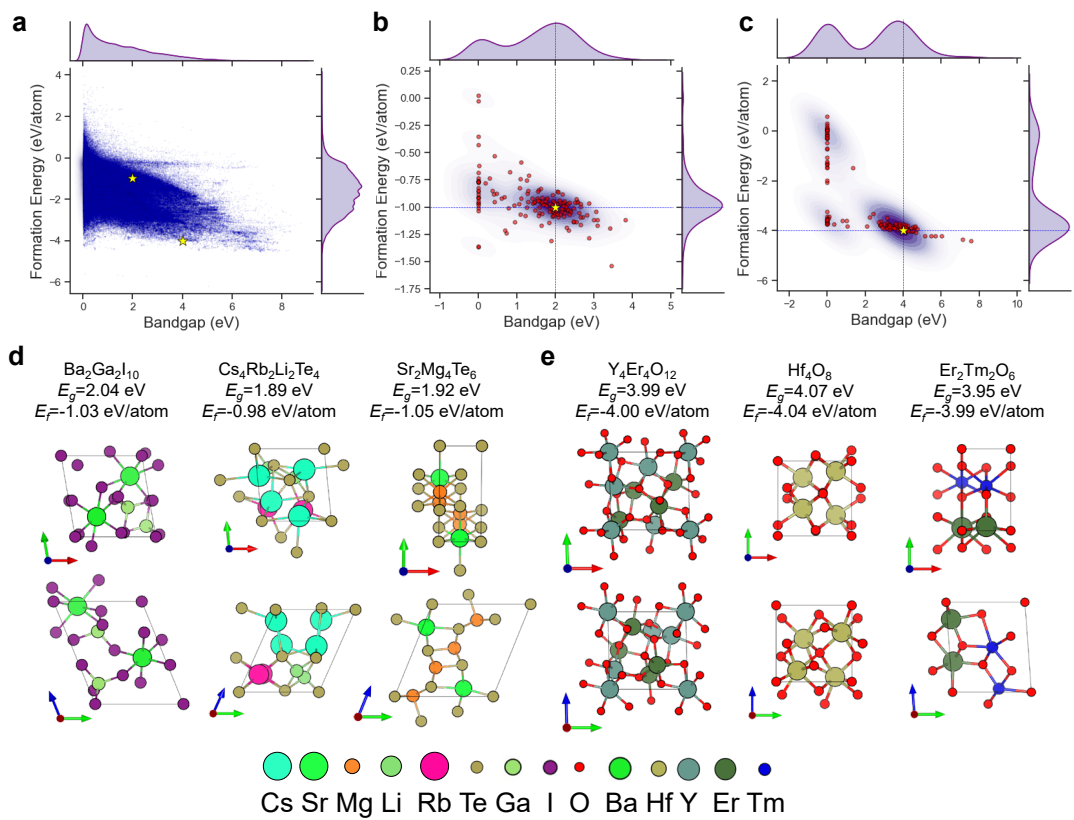

**Figure 6. De novo crystal generation by MatterGPT targeting desired ($E_g$, $E_f$) pairs. a,** training dataset from Alex-20. **b-c,** distribution of formation energy and energy band gap of MatterGPT sampled novel crystals targeting (-1.0, 2.0) and (-4.0, 4.0), respectively. The colour represents kernel density estimate (KDE) plot. **d-e**, top and side views of 3 novel de novo generated crystals with desired ($E_g$, $E_f$).

## CONCLUSIONS AND OUTLOOKS

Inverse design of solid-state materials with desired properties has long been a challenge in materials science. While MatterGPT represents a significant step towards this goal, it still faces challenges due to limited training data. Unlike SMILES-based models that benefit from millions of molecular structures, crystal databases such as the Materials Project [44], JARVIS [44], and Alexandria [33] typically contain only tens of thousands of entries. This data scarcity currently constrains the full potential of MatterGPT, highlighting the critical need for expanding crystal structure databases to enhance inverse design capabilities.

Despite these limitations, MatterGPT demonstrates remarkable versatility. It can be trained for specific tasks such as high-temperature superconductivity [45, 46], energy-related materials [38], etc. The scalability of Transformer-based architectures also allows for expansion to more generalized tasks. Future developments could include concurrent SLICES sequence regression and generation [47], multi-task functionality through task description tokens [48], and integration of reasoning, discovery, and generative capabilities via multimodal LLMs [49]. These advancements could further enhance MatterGPT's utility in crystal design and materials discovery.

MatterGPT demonstrates powerful on-demand crystal generation capabilities, excelling in generating structures with specific attributes and optimizing for multiple properties simultaneously. As a fully open-source project, we provide the complete toolkit including the model architecture, model weights, trained parameters, training data, data cleaning code, training scripts, and post-processing visualization tools. This comprehensive release serves as an open playground for the materials science community, aiming to accelerate collaborative progress in computational materials discovery and design.

## MATERIALS AND METHODS

### SLICES representation for crystals

SLICES is an invertible, invariant crystal representation that encodes three key aspects of crystal information: atomic symbols, node indices, and edge labels[32]. As illustrated in **Figure 1**, SLICES strings incorporate space symbols within the sequence, facilitating a straightforward tokenization process by simply splitting the string at these space symbols. This approach yields tokens that serve as the input for training the MatterGPT model.

Encoding crystal structures to SLICES strings involves: parsing the structure file with Pymatgen [50], constructing a structure graph using the crystal near-neighbor (crystalnn) algorithm [51], and extracting chemical composition, bonding connectivity, and translation vectors to generate the SLICES string.

Decoding SLICES strings to crystal structures uses the SLI2Cry algorithm[32], which comprises three steps: (1) generating an initial structure using Eon's topology-based method, (2) optimizing based on geometry predicted by a modified polarizable force field, and (3) refining the structure with the universal deep learning force field. This process successfully reconstructed 94.95% of original crystal structures, demonstrating SLICES' high invertibility while maintaining invariance to translation, rotation, and permutation.

### Curated datasets

The Alex-20 dataset is derived from the Alexandria database [33, 52], initially comprising materials with less than 20 atoms per unit cell. We applied several filtering steps to ensure the quality and processability of the data:

1. Elemental composition: Due to the limitations of the modified GFN-FF [53] in step (II) of our SLI2Cry reconstruction scheme, we excluded crystals containing atoms with atomic numbers exceeding 86 [32].
2. Structural dimensionality: We removed low-dimensional (0D, 1D, or 2D) structures, as Eon's method used in step (I) of our SLI2Cry reconstruction scheme is not applicable to the fragmented quotient graphs of such materials [54].
3. Metallic exclusion: To focus on semiconducting and insulating materials, we excluded all structures classified as metallic based on their electronic properties to avoid biasing the regression model [55, 56].

After applying these filtering criteria, the final Alex-20 dataset used in this study consists of 280,033 unique crystal structures. This curated dataset ensures that all included materials are suitable for our analysis and can be accurately processed by our computational methods. For the Alex-20 dataset, the SLICES vocabulary comprises 132 tokens in total: 83 atomic symbols, 20 node indices, 27 edge labels, and 2 special tokens ('>' for start and '<' for padding). This comprehensive tokenization scheme enables MatterGPT to effectively learn and generate diverse crystal structures.

## MatterGPT architecture

The MatterGPT model is fundamentally based on the Generative Pre-Training Transformer (GPT) architecture,[19] composed of $n$ blocks. Each block is built upon the Transformer-decoder structure, which consists of a masked self-attention layer followed by a fully connected feedforward layer. The self-attention mechanism is calculated using the "Scaled Dot Product Attention",[17] given by

$$\text{Attention}(\boldsymbol{Q}, \boldsymbol{K}, \boldsymbol{V}) = \text{softmax}\left(\frac{\boldsymbol{Q}\boldsymbol{K}^T}{\sqrt{d}}\right)\boldsymbol{V}$$

where $\boldsymbol{Q}$, $\boldsymbol{K}$, and $\boldsymbol{V}$ are the query, key and value matrixes, respectively, all derived from the same input in the self-attention mechanism. For cross-attention, $\boldsymbol{Q}$ is derived from the input, while $\boldsymbol{K}$ and $\boldsymbol{V}$ are obtained from a different source. The masked self-attention ensures that each token in the sequence attends only to previous tokens and not to future ones, thereby preserving the autoregressive nature of the model.

Moreover, the MatterGPT model leverages a multi-head attention mechanism, where multiple attention heads are employed to capture different aspects of the relationships between tokens. Each head applies the attention function in parallel with its own set of learned linear projections for $\boldsymbol{Q}$, $\boldsymbol{K}$, and $\boldsymbol{V}$. The outputs from these attention heads are then concatenated and linearly transformed to produce the final attention output, enhancing the model's capacity to capture diverse features and dependencies across the sequence.

During the sampling process, to increase the diversity and creativity of MatterGPT, we employ Gumbel-Softmax sampling, which introduces controlled noise into the predicted logit distributions by using a sampling temperature $T>1.0$[57]. In our implementation, we set $T=1.2$. In addition, we utilize nucleus sampling (or top-p sampling) to further enhance sampling diversity while maintaining fluency and coherence [58]. This involves sampling tokens from the top of the probability distribution, defined by a cumulative probability threshold top_p. In our implementation, we set top_p=0.9, ensuring that only tokens within the top 90% of cumulative probability are considered during generation.

Inherited from the Transformer model, MatterGPT is designed to be highly scalable and adaptable, making it suitable for a wide range of tasks within the domain of material science. Its architecture allows for the handling of varying dataset sizes and the ability to adapt to different crystal design challenges with minimal modifications.

## Implementations of MatterGPT

All MatterGPT models are trained for 50 epochs using the Adam optimization algorithm, with an initial learning rate set at 0.0001. The training is conducted with a batch size of 60, an embedding size of 768, a total of 12 Transformer-decoder layers, and 12 attention heads per layer, resulting in approximately 80 million trainable hyperparameters. All the hyperparameters, including learning rate, batch size, and number of layers, are meticulously tuned through a grid search to ensure optimal performance.

The model was implemented in PyTorch [59]. Open-source codes from MolGPT[25] and miniGPT (https://github.com/karpathy/minGPT) were modified to implement MatterGPT. All the codes and data in this work are available on GitHub (https://github.com/xiaohang007/SLICES/tree/main/MatterGPT/). We provide tutorials in Jupyter notebook format, combining code with detailed explanations. The entire runtime environment, including the SLICES v2.0.0 package and all necessary dependencies, is encapsulated in a Docker image. This setup ensures easy result reproduction and facilitates further development. All the training and sampling of MatterGPT are performed on an NVIDIA 4090 GPU.

Table S1 The hyperparameters of MatterGPT model.

| Optimizer | AdamW |
|---|---|
| Learning rate | 0.0001 |
| Batch size | 60 |
| Embedding dropout | 0.1 |
| Attention dropout | 0.1 |
| Output projection dropout | 0.1 |
| Embedding size | 768 |
| Feedforward Size | 3072 |
| Attention heads | 12 |
| Number of layers | 12 |

## Evaluation criteria

The performance of crystal generative models is evaluated using four criteria including validity, uniqueness, novelty and MAPE. The validity of SLICES generated by MatterGPT measures the percentage of SLICES that satisfies all grammar of SLICES representation. The uniqueness of SLICES generated by MatterGPT is measured as the percentage of unique samples out of the total number of valid samples, which is determined by first converting the samples into canonical SLICES strings and then removing duplicates. The higher this measure, the better capability the model can generate diverse samples. The novelty of SLICES generated by MatterGPT measures the percentage of the generated samples out of unique samples are new samples that do not exist in the training dataset. Novelty is computed by comparing the atomic arrangement between every pair of structures that have the same composition via the StructureMatcher utility as implemented in Pymatgen package[50], with the default parameters: ltol=0.2, stol=0.3, angle tol=5.

To characterize the quality of directionality, we calculate mean absolute percentage error (MAPE),

$$\text{MAPE} = \frac{1}{n}\sum_{i=1}^{n}\left|\frac{P_i - T}{T}\right| \times 100$$

where n represents the number of samples, $P_i$ represents material properties of MatterGPT-generated crystals calculated by DFT. $T$ represents the target value.

## DFT calculation workflow

We adopt the just-in-time job management framework custodian, as implemented in Pymatgen, to manage various tasks, including error checking, job management and error recovery[50]. All the DFT calculations were carried out with spin polarization using Vienna Ab initio simulation package (VASP, version=6.3.1) [60]. The projector augmented wave (PAW) method [61] with Perdew, Burke, and Ernzerhof (PBE) parametrization was adopted for the exchange-correlation functional[62]. All the calculation parameters were chosen to be compatible with the MP database[33, 44]. Prior to the formation energy and band gap calculations, all the structures were first relaxed without any symmetry constraints with the MPRelaxSet settings as implemented in Pymatgen[50].

For formation energy calculations, after structural relaxation, the total energy of each compound was obtained, and the formation energy per atom was calculated using the following formula:

$$E_f = \frac{E_{Total} - \sum n_i E_i}{\sum n_i}$$

where $E_{Total}$ is the total energy of the compound, $E_i$ is the energy of the element in its standard state, and $n_i$ represents the number of atoms of element $i$ in the compound.

For energy band gap calculations, following structural relaxation, we performed self-consistent field (SCF) calculations using MPStaticSet to obtain the ground-state electron density. The band structure was then computed along high-symmetry points in the Brillouin zone using MPNonSCFSet, as per Pymatgen's standard settings[50]. The band gap was determined by the energy difference between the conduction band minimum (CBM) and the valence band maximum (VBM). The band structure plots were generated with Pymatgen's standard tools, providing a detailed view of the electronic properties.

## GUI Demo

A simple webapp for inference with trained MatterGPT models is publicly available via HuggingFace spaces at https://huggingface.co/spaces/xiaohang07/MatterGPT_CPU. The app was built using Gradio SDK[63], with crystal visualization and basic information analysis performed using the Atomic Simulation Environment (ASE)[64].

## CRediT authorship contribution statement

Yan Chen: Conceptualization, Methodology, Investigation, Visualization, Writing – original draft, Writing – review & editing, Funding acquisition, Supervision, Writing – review & editing. Hang Xiao: Conceptualization, Methodology, Writing – review & editing, Funding acquisition, Supervision, Funding acquisition, Supervision, Writing – review & editing.

## Declaration of competing interest

The authors declare that they have no known competing financial interests or personal relationships that could have appeared to influence the work reported in this paper.

## Acknowledgments

Yan Chen acknowledges the financial support of National Natural Science Foundation of China (12302140), the Fundamental Research Funds for the Central Universities of China (sxzy012023213), China Postdoctoral Science Foundation (2023M732794) and Postdoctoral Fellowship Program (Grade B) of China Postdoctoral Science Foundation (GZB20230575). H.X. acknowledges the support from the 6th Young Elite Scientist Sponsorship Program by China Association for Science and Technology (Grant No. 2020QNRC001), National Natural Science Foundation of China (Grant No. 22203066).